\documentclass[fleqn,usenatbib]{mnras}

\usepackage{newtxtext,newtxmath}
\usepackage{tablefootnote}
\usepackage[T1]{fontenc}
\usepackage{ae,aecompl}

\usepackage{graphicx}	% Including figure files
\usepackage{amsmath}	% Advanced maths commands
\usepackage{amssymb}	% Extra maths symbols
\usepackage{overpic}
\newcommand{\epse}{\epsilon_{e}}
\newcommand{\epsB}{\epsilon_{B}}

\newcommand{\thetaobs}{\theta_v}
\newcommand{\Mej}{M_{\mathrm{ej}}}
\newcommand{\umin}{u_{\mathrm{min}}}
\newcommand{\umax}{u_{\mathrm{max}}}
\newcommand{\Einj}{E_{\mathrm{inj}}}

\title[The outflow of GW 170817]{The outflow structure of GW170817 from late time broadband   observations}

\author[Troja et al.]{
E. Troja,$^{1,2}$\thanks{E-mail: eleonora.troja@nasa.gov}
L. Piro,$^{3}$
G. Ryan,$^{1}$
H. van Eerten,$^{4}$
R. Ricci,$^{5}$ 
M. H.~Wieringa,$^{6}$
\newauthor
S. Lotti,$^{3}$
T. Sakamoto,$^{7}$ 
and S. B. Cenko$^{2}$
\\
$^{1}$ Department of Astronomy, University of Maryland, College Park, MD 20742-4111, USA \\
$^{2}$Astrophysics Science 
Division, NASA Goddard Space Flight Center, 8800 Greenbelt Rd, Greenbelt, MD 20771, USA\\
$^{3}$INAF, Istituto di Astrofisica e Planetologia Spaziali, via Fosso del Cavaliere 100, 00133 Rome, Italy\\
$^{4}$Department of Physics, University of Bath, Claverton Down, Bath BA2 7AY, United Kingdom\\
$^{5}$INAF-Istituto di Radioastronomia, Via Gobetti 101, I-40129, Italy\\
$^{6}$CSIRO Astronomy and Space Science, P.O. Box 76, Epping NSW 1710, Australia\\
$^{7}$%Department of Physics and Mathematics, 
Aoyama Gakuin University, 5-10-1 Fuchinobe, Chuoku, Sagamiharashi Kanagawa 252-5258, Japan
}

\date{Accepted: April 4, 2018}

\pubyear{2018}

\begin{document}
\label{firstpage}
\pagerange{\pageref{firstpage}--\pageref{lastpage}}
\maketitle

% Abstract of the paper
\begin{abstract}
We present our broadband study of GW170817 from radio to hard X-rays, including {\it Chandra} and {\it NuSTAR} 
observations, and a multi-messenger analysis including LIGO constraints. The data are compared with predictions from a wide range of models, providing the first detailed comparison between non-trivial cocoon and jet models. Homogeneous and power-law shaped jets, as well as simple cocoon models are ruled out by the data, while both a Gaussian shaped jet and a cocoon with energy injection can describe the current dataset for a reasonable range of physical parameters, consistent with the typical values derived from short GRB afterglows.
We propose that these models can be unambiguously discriminated by future observations measuring the post-peak behaviour, with $F_\nu \propto t^{\sim -1.0}$ for the cocoon and $F_\nu \propto t^{\sim -2.5}$ for the jet model.
\end{abstract}

% Select between one and six entries from the list of approved keywords.
% Don't make up new ones.

%%%%%%%%%%%%%%%%%%%%%%%%%%%%%%%%%%%%%%%%%%%%%%%%%%

%%%%%%%%%%%%%%%%% BODY OF PAPER %%%%%%%%%%%%%%%%%%

\section{Introduction}
The discovery of GW170817 and its electromagnetic counterparts (GRB170817A and AT2017gfo; \citealt{gbm17,coulter17,mma2017}) ushered in a new era of multi-messenger astrophysics, in which both gravitational waves and photons provide complementary views of the same source. 
While observations at optical and infrared wavelengths unveiled the onset and evolution of a radioactive-powered transient, known as kilonova, observations at X-rays and, later, radio wavelengths probed a different component of emission, likely originated by a relativistic outflow launched by the merger remnant. 
\cite{Troja2017} explained the observed X-ray and radio data as the onset of a standard short GRB (sGRB) afterglow viewed at an angle (off-axis).  
However, as already noted in \cite{Troja2017} and \cite{Kasliwal2017}, a standard top-hat jet model could explain the afterglow dataset collected at early times, but failed to account for the observed gamma-ray emission. 
Based on this evidence, \cite{Troja2017} suggested that a structured jet model \citep[e.g.][]{zhang04,kathi18} provided a coherent description of the entire broadband dataset.
Within this framework, the peculiar properties of GRB170817A/AT2017gfo could be explained, at least in part, by its viewing angle (see also \citealt{Lazzati2017,LambKobayashi2017}).
An alternative set of models invoked the ejection of a mildly relativistic wide-angle outflow, either a jet-less fireball \citep{salafia17} or a cocoon \citep{Nagakura2014, Hallinan2017}. In the latter scenario, the jet might be chocked by the merger ejecta \citep{Mooley2017}, and the observed gamma-rays and broadband afterglow emission are produced by the expanding cocoon. The cocoon may be energized throughout its expansion by continuous energy injection.
In this paper   detailed models of structured jet and cocoon, from its simplest to more elaborate version, are compared with the latest radio to X-ray data. Predictions on the late time evolution are derived, and a unambiguous  measurement capable of disentangling the outflow geometry, jet vs cocoon, is presented.

\section{Observations}

\subsection{X-rays}

The  Chandra X-ray Observatory and the Nuclear Spectroscopic Telescope ARray (NuSTAR) re-observed the field of GW170817 soon after the target came out of sunblock.  Chandra data were reduced and analyzed in a standard fashion using CIAO v.~4.9 and CALDB~4.7.6. 
The NuSTAR data were reduced using standard settings of the pipeline within the latest version of NuSTAR Data Analysis Software.  
Spectral fits were performed with XSPEC by minimizing the Cash statistics. 
A log of observations and their results is reported in Table~1. 

We found that the X-ray flux at 108~d is $\approx$5 times brighter than earlier measurements taken in August 2017 (Figure~1), thus confirming our previous findings of a slowly rising X-ray afterglow \citep{Troja2017}. 
By describing the temporal behavior with a simple power-law function, we derive an index $\alpha \approx$ 0.8, consistent with the constraints from radio observations \citep{Mooley2017}. During each observations, no significant temporal variability is detected on timescales of $\approx$1~d or shorter. 

In order to probe any possible spectral evolution we computed the hardness ratio \citep{Park2006}, defined as HR=H-S/H+S,
where H are the net source counts in the hard band (2.0-7.0 keV) and S are the net source counts in the soft band (0.5-2.0 keV). This revealed a possible softening of the spectrum, with HR$\approx$-0.1 for the earlier observations ($\lesssim$ 15 d) and HR$\approx$-0.5 for the latest observations ($>$100~d). However, the large statistical uncertainties prevent any firm conclusion.

\begin{figure}
\includegraphics[width=\columnwidth]{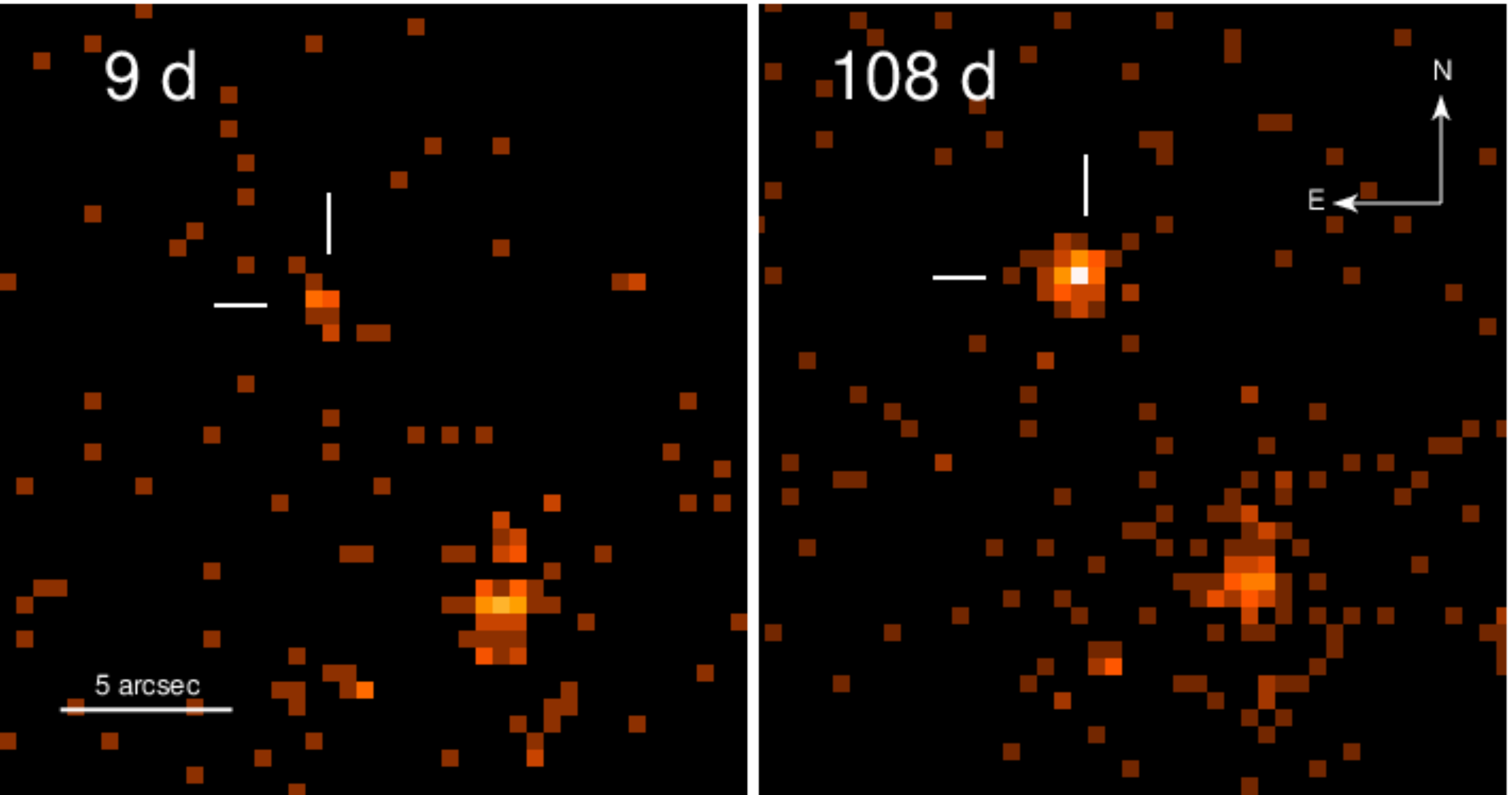}
     \caption{Exposure corrected X-ray images of the field of GW 170817 taken at 9 days (left; \citealt{Troja2017}) and 108 days (right) after the merger. 
The X-ray counterpart, marked by the crossed lines, significantly brightened between the two epochs.}
     \label{fig:image}
     \vspace{-0.3cm}
\end{figure}
    
\subsection{Radio}

The target GW170817 was observed with the Australia Telescope Compact Array in three further epochs after Sept 2017. 
Observations were carried out at the center frequencies of 5.5 and 9 GHz with a bandwidth of 2 GHz.
For these runs the bandpass calibrator was 0823-500, the flux density absolute scale was determined using 1934-638 and the source 1245-197 was used as phase calibrator. Standard MIRIAD procedures were used for loading, inspecting, flagging, calibrating and imaging the data. The target was clearly detected at all epochs, our measurements are reported in Table~1. 

\begin{figure}
\includegraphics[width=0.9\columnwidth]{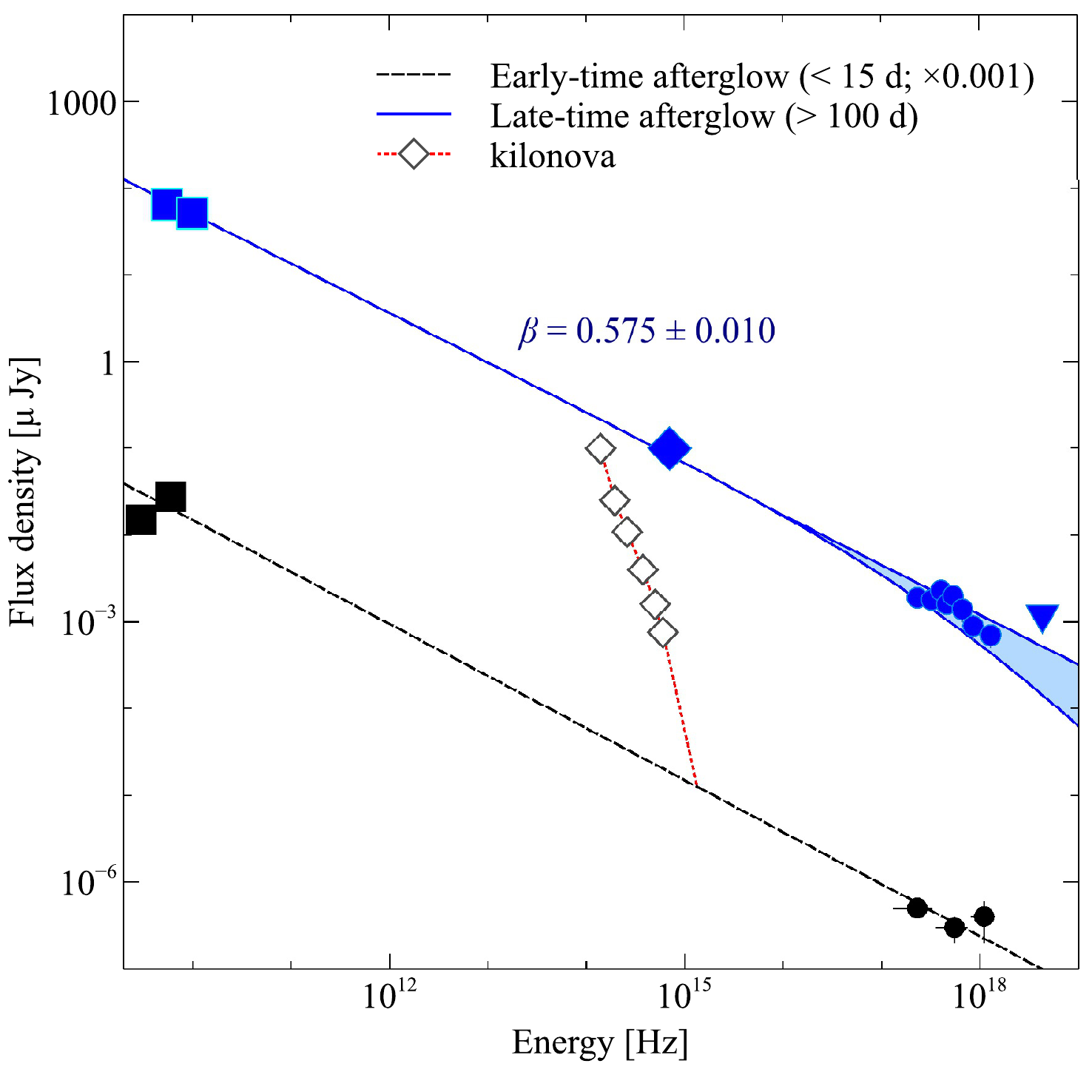}
     \caption{Spectral energy distribution of GW170817
     at early (black) and late (blue) times.
     Optical data are from \citet{Troja2017,Lyman2018},
     early radio data are from \citet{Hallinan2017}.}
     \label{fig:sedx}
     \vspace{-0.3cm}
\end{figure}

The broadband spectrum, from radio to  optical \citep{Lyman2018} to hard X-rays (Figure~\ref{fig:sedx}), can be fit with a simple power-law model with spectral index $\beta$=0.575$\pm$0.010 and no intrinsic absorption. A fit with a realistic afterglow spectrum \citep{GranotSari2002} constrains the cooling break to $\nu_c \gtrsim$1~keV (90\% confidence level). 

\begin{table}
\begin{center}
 	\caption{X-ray and radio observations of GW170817. The quoted uncertainties are at the 68\% confidence level.}
 	\label{tab:obs}	      
 	\begin{tabular}{ccccccc}
    	\hline
        T-T$_0$ & Exposure & $\beta$  & Flux$^1$
 & Energy \\
        \hline
       8.9   &  49.4 ks &  -0.1$\pm$0.5  & 4.0$\pm$1.1 & 0.3-10 keV \\ 
       15.2  &  46.7 ks &  0.6$\pm$0.4 & 5.0$\pm$1.0 & 0.3-10 keV \\
       103   &  70.7 ks &  0.6          & $<$25$^2$ & 3-10 keV. \\
             &          &  0.6          & $<$60$^2$  & 10-30 keV\\       
       107.5 &  74.1 ks &  0.6$\pm$0.2 & 26$\pm$3  & 0.3-10 keV \\
       110.9 &  24.7 ks &  0.9$\pm$0.4 & 23$\pm$4  & 0.3-10 keV\\
       158.5 &  104.9 ks & 0.67$\pm$0.12 & 26$\pm$2 & 0.3-10 keV\\
       \hline
       75   &  12 hrs &   -- & 50$\pm$8 & 5.5 GHz \\
            &         &      & 31$\pm$5 & 9 GHz \\ 
       92   &   9 hrs &  -- & 49$\pm$8 & 5.5 GHz \\
            &         &      & 21$\pm$7 & 9 GHz \\
      107   &    12 hrs    &    --   & 65$\pm$8 & 5.5 GHz \\
             &         &      & 52$\pm$7 & 9 GHz \\
        \hline
    \end{tabular}\\
    \end{center}
    {$^1$ Units are 10$^{-15}$\,erg\,cm$^{-2}$\,s$^{-1}$ for X-ray fluxes, and $\mu$Jy for radio fluxes.}
    {$^2$ NuSTAR 3$\sigma$ upper limit.} 
    \end{table}

\section{Ejecta and Afterglow Modeling}

\label{se:model}

\subsection{Jet and cocoon}

The standard model for sGRB afterglows describes these in terms of synchrotron emission from a decelerating and decollimating relativistic jet. More recent studies have argued for the additional presence of a slower moving cocoon (e.g. \citealt{Nagakura2014,Lazzati2017,Mooley2017}) also in the case of sGRBs. Numerical studies of jet breakouts have revealed a range of possibilities for jet velocities and initial angular structure. In the case of jetted outflow seen from a substantial angle $\theta_\nu$ (i.e. larger than the opening angle $\theta_c$ of a top-hat flow, or in the wings of a jet with energy dropping as a function of angle), relativistic beaming effects will delay the observed rise time of the jet emission.

The dynamics of the jet component can be treated semi-analytically at various levels of detail (e.g. \citealt{RossiLazzatiRees2002, Dalessio2006}), depending on additional assumptions for the angular structure of the jet. Early X-ray and radio observations already rule out \citep{Troja2017} the universal jet structure (i.e. a power-law drop in energy at larger angles), and we will not discuss this option further here. For the \emph{Gaussian} structured jet, we assume energy drops according to $E(\theta) = E_0 \exp[-\theta^2 / 2 \theta^2_c]$, up to a truncating angle $\theta_w$. We approximate the radial jet structure by a thin homogeneous shell behind the shock front. Top hat and Gaussian jet spreading are approximated following the semi-analytical model from \cite{vanEertenZhangMacFadyen2010}, with the Gaussian jet implemented as a series of concentric top hat jets. This spreading approximation was tuned to simulation output \citep{vanEertenZhangMacFadyen2010} that starts from top-hat initial conditions but develops a more complex angular structure over time. Since the off-axis jet emission will fully come into view after deceleration, deceleration radius and initial Lorentz factor value no longer impact the emission and are not included in the model.

The cocoon is similarly treated using a decelerating shell model, now assuming sphericity and including pre-deceleration stage. For a direct comparison to \cite{Mooley2017}, we have added a mass profile that accounts for velocity stratification in the ejecta and thereby provides for ongoing energy injection. The total amount of energy in the slower ejecta above a particular four-velocity is assumed to be a power-law $E_{>u}(u) = E_{inj}u^{-k}$ for $u \in [\umin,\umax]$ (note that for relativistic flow $u \to \gamma$). The energy from a slower shell is added to the forward shock once this reaches the same velocity. The total cocoon energy (and therefore light curve turnover time) is thus dictated by $\umin$. We assume an initial cocoon mass of $M_{ej}$.
Both jet and cocoon emerge into a homogeneous environment with number density $n$.

We estimate the emission of the ejecta using a synchrotron model \citep{SariPiranNarayan1998}. Electrons are assumed to be accelerated to a power law distribution in energy of slope $-p$, containing a fraction $\epsilon_e$ of the post-shock internal energy. A further fraction $\epsilon_B$ is estimated to reside in shock-generated magnetic field energy. We integrate over emission angles to compute the observed flux for an observer at luminosity distance $d_L$ and redshift $z$.

The key distinguishing features of the various models are their rise slope, peak time and decay slope. Above synchrotron injection break $\nu_m$, a cocoon will show a steeply rising flux $F_\nu \propto t^{\sim 3}$. A top-hat jet will show a steeper rise while a Gaussian energy profile will show a more gradual rise than a top-hat jet (see Extended Fig. 3 of \citealt{Troja2017}). Gaussian and top-hat jets will have peak times following \citep{Troja2017}:
\begin{equation}
     t_{peak}\propto\left(\frac{E_{0,50}}{n_{-3}}\right)^{1/3} \left(\theta_v-\theta_c\right)^{2.5} \mathrm{days}, \qquad \mathrm{(jet)}
 	\label{eq:jet_peak}
 \end{equation}
while a cocoon outflow whose energy is dominated by the slow ejecta will peak at a time according to 
\begin{equation}
     t_{peak} \approx 81 \left(\frac{k E_{50}}{n_{-3}\ \umin^8}\right)^{1/3} \mathrm{days}.  \qquad \mathrm{(cocoon)}
 	\label{eq:cocoon_peak}
 \end{equation}
Here $E_{0,50}$ is the on-axis equivalent isotropic energy in units to $10^{50}$ erg, $E_{50}$ the total cocoon energy following energy injection in the same units, and $n_{-3}$ circumburst density in units of $10^{-3}$ cm$^{-3}$.

\begin{figure}
\includegraphics[width=\columnwidth]{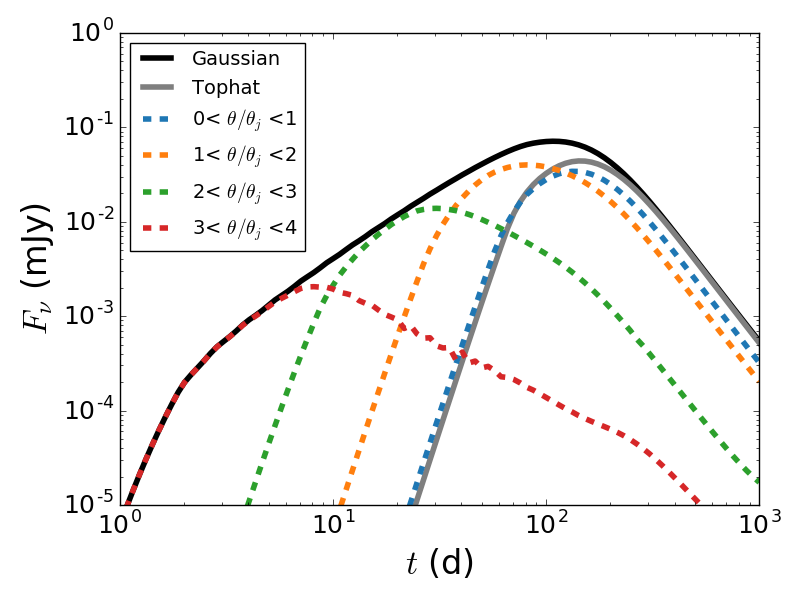}
     \vspace{-0.3cm}
\caption{Emission at $6$ GHz from Gaussian (solid black) and top-hat (solid grey) jets viewed off-axis. The emission from the Gaussian is further divided by the polar angle from which the emission originated (dashed colored), separated by multiples of $\theta_c$.  Emission from the ``wings'' ($\theta \gg \theta_c$) rises first and dominated the early emission.  Inner regions dominate progressively later, leading to a slow rise in observed emission.  Once the entire jet is in view emission peaks and decays following a top-hat profile.  
     }
     \label{fig:jetStruct}
\end{figure}

Cocoon and jet models differ in their expected post-peak downturn slopes. For the cocoon, $\sim t^{-1.0}$ (a decelerating spherical fireball) is expected, while for jet models $\sim t^{-2.5}$ is expected due to a combination of jet spreading dynamics and the entire jet having come into view (Figure~\ref{fig:jetStruct}). A jet \emph{plus} cocoon might yield an intermediate slope value, while still confirming the collimated nature of the sGRB.

\begin{figure*}
	\includegraphics[width=\columnwidth]{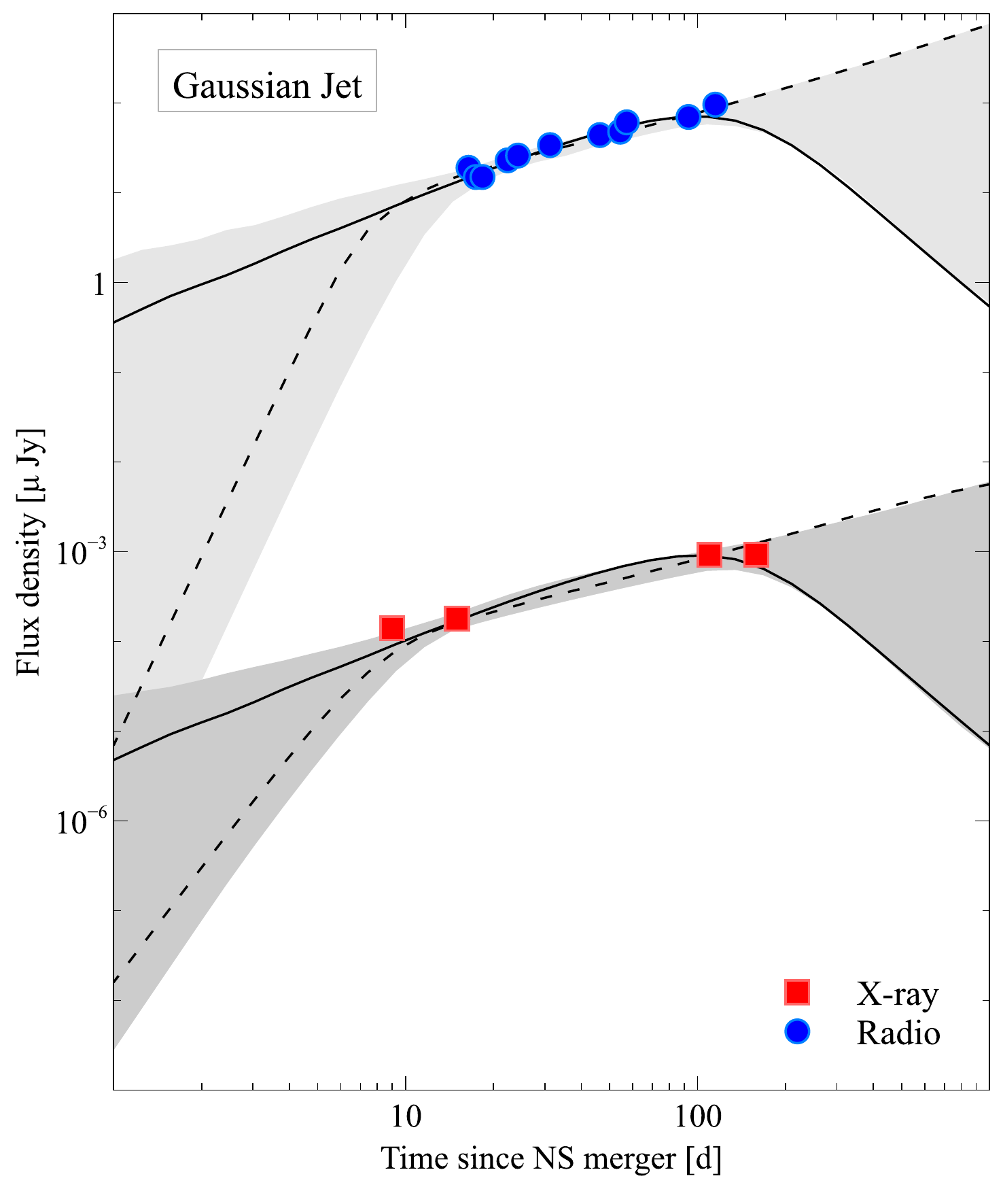}
    \hspace{0.3cm}
    	\includegraphics[width=\columnwidth]{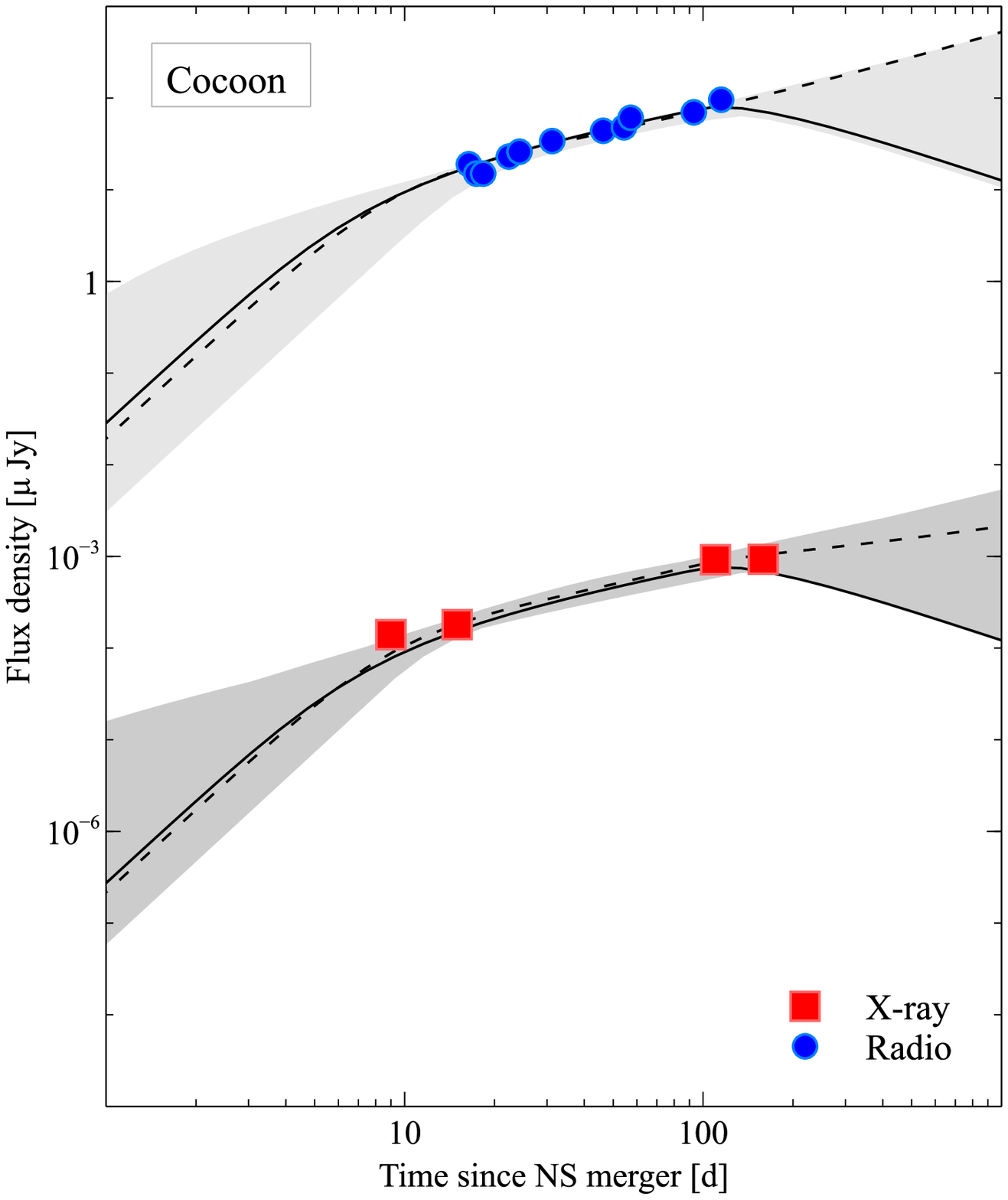}
    \caption{Left panel: The Gaussian structured jet light curve at 3 GHz (upper curves) and 5~keV (lower curves). 
Right panel: The cocoon light curve at 3 GHz and 5 keV. 
 The range of possible flux values attained by models in the top 68\% of the posterior is shaded grey.  Of this set, the light curves with the earliest and the latest peak time are shown by the solid and dashed lines, respectively.}
    \label{fig:lcjet}
\end{figure*}

\subsection{Bayesian model fit including LIGO/VIRGO constraints}

We perform a Bayesian Markov-Chain Monte Carlo (MCMC) model fit to data from synthetic detections generated from our semi-analytical cocoon and jet models. At fixed distance, the off-axis structured jet models considered here are fully determined by the set of eight parameters $\Theta_{\mathrm{jet}} = \{\thetaobs, E_0, \theta_c, \theta_w, n, p, \epse, \epsB\}$. The isotropic cocoon with a power law velocity distribution, on the other hand, requires nine parameters $\Theta_{\mathrm{cocoon}} = \{\umax, \umin, \Einj, k, \Mej, n, p, \epse, \epsB\}$. We generate samples of the posterior for both models using the affine-invariant ensemble MCMC sampler implemented in the {\tt emcee} package (\citealt{GoodmanWeare2010, Foreman-Mackey2013}). For both the cocoon and jet models we initialize the MCMC walkers in a small ball near the maximum of the posterior, calculated through trial runs.  We run each model with 300 walkers for 128000 steps, dropping the first 36000 steps as an initial burn-in phase, generating $\sim 3 \times 10^7$ posterior samples.

We assign independent priors for each parameter, uniform for $\theta_c$, $\theta_w$, $k$, and $p$, and log-uniform for $E_0$, $\umax$, $\umin$, $\Einj$, $\Mej$, $n$, $\epse$, and $\epsB$.  The viewing angle $\thetaobs$ is given a prior $p(\thetaobs) \propto \sin \thetaobs$. This is proportional to the solid angle subtended at this viewing angle and is the expected measured distribution of randomly oriented sources in the sky if observational biases resulting from jet dynamics and beaming are not accounted for. Parameters are given wide bounds so as to ensure the parameter space is fully explored. The era of multi-messenger astronomy allows us to directly link the observational constraints from the different channels. The upper bound on the viewing angle is therefore chosen to include the 95\% confidence interval from the LIGO analysis of GW170817A assuming either value (CMB or SNe) of the Hubble constant \citep{LIGO2017standardsiren}.

The LIGO/VIRGO analysis of GW170817A includes the inclination $i$, the angle between the total angular momentum vector of the binary neutron star system and the line of sight \citep{LIGO2017standardsiren}.  Assuming this angle to be identical to the viewing angle $\thetaobs$, we can incorporate their posterior distributions $p_{\mathrm{GW}}(\thetaobs)$ into our own analysis of the Gaussian structured jet. We do this by applying a weighting factor $p_{\mathrm{GW}}(\thetaobs) / p(\thetaobs)$, where $p(\thetaobs)$ is our prior, to the MCMC samples.  This is valid so long as the MCMC adequately samples the region of parameter space favored by $p_{\mathrm{GW}}$, which we confirmed for our analysis.  

LIGO/VIRGO report three distributions $p_{\mathrm{GW}}$, one using only the gravitational wave data and two incorporating the known redshift of host galaxy NGC 4993, utilizing the value of the Hubble constant reported by either the Planck collaboration or the SHoES collaboration (\citealt{Planck2016, Riess2016, LIGO2017standardsiren}). We incorporate the latter two distributions into our analysis.  The reported distribution functions (Fig 3 of \citealt{LIGO2017standardsiren}) were digitized and found to be very well fit by Gaussian distributions in $\cos \thetaobs$.  We take the distributions to be:
\begin{equation}
	p_{\mathrm{GW}}(\thetaobs) \propto \exp\left[-\frac{1}{2}\left(\frac{\cos \thetaobs - \mu_0}{\sigma}\right)^2\right]\ ,
\end{equation}
where for the Planck $H_0$ $\mu_0=0.985$ and $\sigma = 0.070$ and for the SHoES $H_0$ $\mu_0 = 0.909$ and $\sigma = 0.068$. The overall normalization of $p_{\mathrm{GW}}$ need not be specified in this approach.

\section{Results and discussion}

\begin{table*}
 	\centering
 	\caption{Constraints on the Gaussian jet and Cocoon model parameters.  Reported are the median values of each parameter's posterior distribution with symmetric 68\% uncertainties (ie. the 16\% and 84\% quantiles). }
 	\label{tab:MCMC}
    \begin{tabular}{cc}
 	\begin{tabular}{lccc}
 		\hline
        & \multicolumn{1}{c}{Jet}& \multicolumn{1}{c}{Jet+GW+Planck} & \multicolumn{1}{c}{Jet+GW+SHoES} \\
        \hline
        Parameter & Med. & Med.  & Med. \\
        \hline
 		$\thetaobs$ & 		$0.51^{+0.20}_{-0.22}$  
        				&	$0.32^{+0.13}_{-0.13}$ 
                        &	$0.43^{+0.13}_{-0.15}$  \\[3pt]
        $\log_{10}E_0$ & 	$52.50^{+1.6}_{-0.79}$ 
        				&	$52.73^{+1.30}_{-0.75}$ 
                        &	$52.52^{+1.4}_{-0.71}$ \\[3pt]
        $\theta_c$ & 		$0.091^{+0.037}_{-0.040}$ 
        				&	$0.057^{+0.025}_{-0.023}$ 
                        &	$0.076^{+0.026}_{-0.027}$ \\[3pt]
        $\theta_w$ & 		$0.55^{+0.65}_{-0.22}$  
        				&	$0.62^{+0.65}_{-0.37}$ 
                        &	$0.53^{+0.70}_{-0.24}$ \\[3pt]
                        \\
        \hline
 		$\log_{10} n_0$ & 	$-3.1^{+1.0}_{-1.4}$  
        				&	$-3.8^{+1.0}_{-1.3}$ 
                        &	$-3.24^{+0.91}_{-1.3}$  \\[3pt]
        $p$ & 				$2.155^{+0.015}_{-0.014}$ 
        				&	$2.155^{+0.015}_{-0.014}$ 
                        &	$2.155^{+0.015}_{-0.014}$  \\[3pt]
        $\log_{10}\epse$ & 	$-1.22^{+0.45}_{-0.80}$  
        				&	$-1.51^{+0.53}_{-0.89}$ 
                        &	$-1.31^{+0.46}_{-0.78}$  \\[3pt]
        $\log_{10}\epsB$ & 	$-3.38^{+0.81}_{-0.45}$  
        				&	$-3.20^{+0.92}_{-0.58}$ 
                        &	$-3.33^{+0.82}_{-0.49}$ \\
        \hline
 		$\log_{10} E_{tot}$ & $50.26^{+1.7}_{-0.69}$  
        				&	$50.16^{+1.1}_{-0.67}$ 
                        &	$50.19^{+1.41}_{-0.65}$ \\
 		\hline
 	\end{tabular} & \begin{tabular}{lc}
 		\hline
        & \multicolumn{1}{c}{Cocoon} \\
        \hline
        Parameter & Med. \\
        \hline
 		$\log_{10} \umax$ & $0.93^{+0.34}_{-0.36}$  \\[3pt]
        $\log_{10} \umin$ & $-2.2^{+1.9}_{-1.9}$    \\[3pt]
        $\log_{10} \Einj$ & $54.7^{+1.6}_{-2.7}$  \\[3pt]
        $k$ 			  & $5.62^{+0.93}_{-1.1}$  \\[3pt]
        $\log_{10} \Mej$ & $-7.6^{+2.1}_{-1.7}$  \\
        \hline
 		$\log_{10} n_0$ & 	$-5.2^{+2.2}_{-2.0}$  \\[3pt]
        $p$ & 				$2.156^{+0.014}_{-0.014}$ \\[3pt]
        $\log_{10}\epse$ & 	$-1.33^{+0.93}_{-1.3}$  \\[3pt]
        $\log_{10}\epsB$ & 	$-2.5^{+1.5}_{-1.1}$  \\
        \hline
 		$\log_{10} E_{tot}^*$ & $52.84^{+0.97}_{-1.3}$  \\
 		\hline
 	\end{tabular}
    \end{tabular}
\end{table*}

\subsection{Constraints to Jet and Cocoon Models}

The slow rise $\propto t^{0.8}$ observed in the radio and X-ray data is inconsistent with the steep rise predicted by the basic cocoon model without energy injection \citep{Hallinan2017}. The top hat jet model is also characterized by a steep rise and a narrow plateau around the peak, as the result of the jet energy being limited to a narrow core. This is a well defined property of this model, rather independent of the other afterglow parameters, and the late-time observations of GW170817 allow us to robustly reject it.

Some degree of structure in the GRB outflow is therefore implied by the late-time observations. The universal jet model is too broad and over-predicts the early time X-ray flux \citep{Troja2017}.
As discussed in the next section, our models of Gaussian jet and cocoon with energy injection provide equally good fits to the current dataset, and are basically indistinguishable until the peak time. 
However, as discussed in section \ref{se:model}, the post-peak slopes are expected to differ, with $F_\nu \propto t^{\sim -1.0}$ for the cocoon and $F_\nu \propto t^{\sim -2.5}$, and intermediate values indicating a combination of directed outflow plus cocoon. Future observations will be critical to distinguish the nature of the outflow (collimated vs isotropic).

\subsection{Bayesian analysis results}

The MCMC fit results for the Gaussian jet and cocoon models are summarized in Table \ref{tab:MCMC}.  Both models are consistent with current data and have similar quality of fit with $\chi^2$ per degree of freedom near unity.  Corner plots showing the 1D marginalized posterior distributions for each parameter and the 2D marginalized posterior distributions for each pair of parameters is included in the on-line materials.

The Gaussian jet model prefers a narrow core ($\theta_c \sim 0.09$ rad) with wings truncated at several times the width of the core.  The viewing angle is significant ($\thetaobs \sim 0.5$) but degenerate with $\theta_c$, $E_0$, and $n_0$. The large uncertainties on individual parameters is partially a result of these degeneracies within the model. The viewing angle correlates strongly with $\theta_c$ and $n_0$ and anti-correlates with $E_0$. Figure \ref{fig:lcjet} (left panel) shows the range of possible X-ray and radio light curves of the Gaussian structured jet, pulled from the top 68\% of the posterior.  There is still significant freedom within the model, which will be better constrained once the emission peaks. 

Incorporating constraints on $\thetaobs$ from LIGO/VIRGO tightens the constraints on $\theta_c$, $E_0$, and $n_0$ due to the correlations between these parameters.   Using either the Planck or SHoES $H_0$ decreases the likely $\theta_v$ and $\theta_c$ significantly, with comparatively small adjustments to $E_0$ and $n_0$.

The cocoon model strongly favours an outflow of initial maximum four-velocity $u \in [3.7,18.6]$ whose energy is dominated by a distribution of slower ejecta.  The mass of the fast ejecta $\Mej$ and the low-velocity cut-off $\umin$ are very poorly constrained.  The total energy of the outflow is strongly dependent on $\umin$, which can only be determined by observing the time at which the emission peaks.  Figure \ref{fig:lcjet} (right panel) shows the range of possible X-ray and radio light curves of the cocoon model.

The posterior distributions of $\epse$ and $\epsB$ under both models are consistent with theoretical expectations, with the cocoon model providing weaker constraints than the jet.  Both models very tightly constrain $p = 2.156 \pm 0.015$, a consequence of simultaneous radio and x-ray observations, and place the cooling frequency above the X-ray band.  The cocoon model tends to prefer a larger total energy and smaller circumburst density than the Gaussian jet, although the posterior distributions of both quantities are broad.

\subsection{Implications for the prompt $\gamma$-ray emission}

In the case of a Gaussian jet, it is natural to consider the question whether the sGRB and its afterglow would have been classified as typical, had the event been observed on-axis. The afterglow values for $E_0$, $\theta_c$ and $\theta_w$ are indeed consistent with this notion. As discussed previously by \cite{Troja2017}, we expect the observed $\gamma$-ray isotropic equivalent energy release $E_{\gamma, obs} \sim 5 \times 10^{46}$ erg to scale up to a typical value of $E_{\gamma, OA} \sim 2 \times 10^{51}$ erg, once the orientation of the jet is accounted for. This implies a ratio $\thetaobs / \theta_c = \sqrt{2 \ln[ E_{\gamma, OA} / E_{\gamma, obs}]} \approx 4.6$. From our afterglow analysis we infer a value of $5.6\pm 0.9$ ($95\%$ uncertainty) for this ratio, accounting for the correlation between the two angles shown by our fit results. Our inferred range of $\thetaobs / \theta_c$ lies marginally above the typical value, but remains consistent with expected range of sGRB energetics.
%Again, this result is consistent with a typical sGRB.

While the structured jet model implies that GRB~170817A would have been observed to be a typical sGRB when seen on-axis, the cocoon model implies that it belongs to a new class of underluminous gamma-ray transients. In the former case, the origin of standard sGRBs as due to neutron star mergers can be considered confirmed, with the added benefit of being able to combine multi-messenger information about jet orientation into a single comprehensive model fit to the data (Table~2). In the latter case, 
the future detection rate of multi-messenger events is potentially  higher and dominated by these failed sGRBs. 
 Continued monitoring at X-ray, optical and radio wavelengths is important to discriminate between these two different scenarios.

\vspace{-0.2cm}

\section{Conclusions}
 Our modeling of the latest broad band data confirms 
 that a jetted outflow seen off-axis is consistent with the data \citep{Troja2017}. The late-time data favor a Gaussian shaped jet profile, while  homogeneous and a power law jets are ruled out. 
 A simple spherical cocoon model also fails to reproduce the observed behaviour and, to be successful, a cocoon with energy injection from earlier shells catching up with the shock is required.
{ Both models can describe the data for a reasonable range of physical parameters, within the observed range of sGRB afterglows}.
A Gaussian jet and a re-energized cocoon are  presently indistinguishable but we predict a different behaviour in their post-break evolution once the broadband signal begins to decay, with $F_\nu \propto t^{\sim -1.0}$ for the cocoon and $F_\nu \propto t^{\sim -2.5}$, and intermediate values indicating a combination of directed outflow plus cocoon.
%Furthermore in this regime the behaviour of the cooling frequency is also expected to be different, with $.
While the cocoon model invoke a new class of previously unobserved phenomena, the Gaussian jet provides a self-consistent model for both the afterglow and the prompt emission and explains the observed properties of GW170817 with a rather normal sGRB seen off-axis.

%The last numbered section should briefly summarise what has been done, and describe
%the final conclusions which the authors draw from their work.

%\section*{Acknowledgements}

%The Acknowledgements section is not numbered. Here you can thank helpful
%colleagues, acknowledge funding agencies, telescopes and facilities used etc.
%Try to keep it short.

%%%%%%%%%%%%%%%%%%%%%%%%%%%%%%%%%%%%%%%%%%%%%%%%%%

%%%%%%%%%%%%%%%%%%%% REFERENCES %%%%%%%%%%%%%%%%%%

% The best way to enter references is to use BibTeX:

%\bibliographystyle{mnras}
%\bibliography{example} % if your bibtex file is called example.bib

% Alternatively you could enter them by hand, like this:
% This method is tedious and prone to error if you have lots of references

\begin{figure*}
	\includegraphics[scale=0.35]{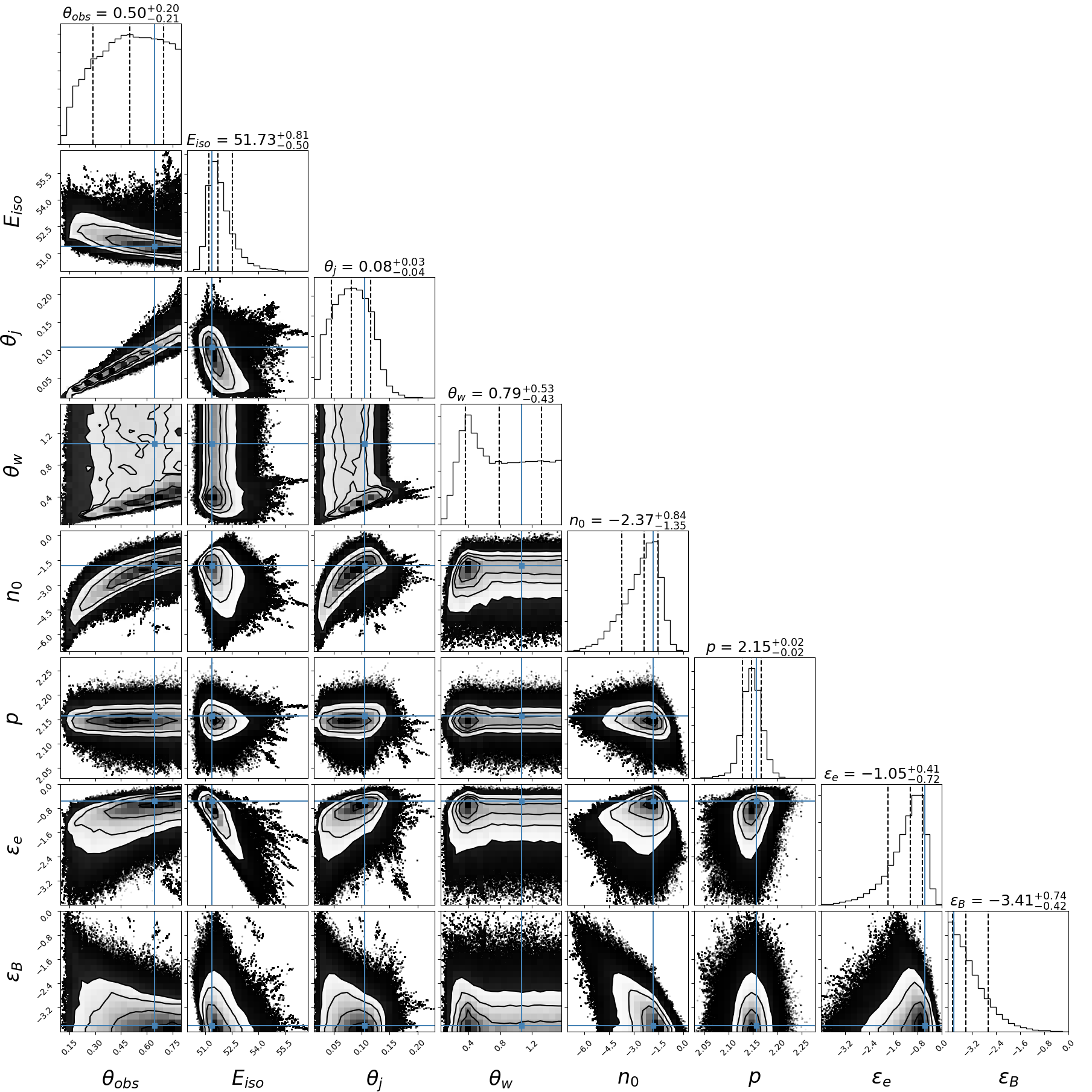}
    \caption{Fit result for the jet model. This ``corner plot'' shows all one-dimensional (diagonal) and two-dimensional (off-diagonal) projections of the posterior pdf. 
    The best-fit value (maximum posterior probability) is shown in blue. Dotted lines mark the 16\%, 50\%, and 84\% quantiles of the marginalized posteriors for each parameter. }
    \label{fig:lcjet}
\end{figure*}

\begin{figure*}
	\includegraphics[scale=0.35]{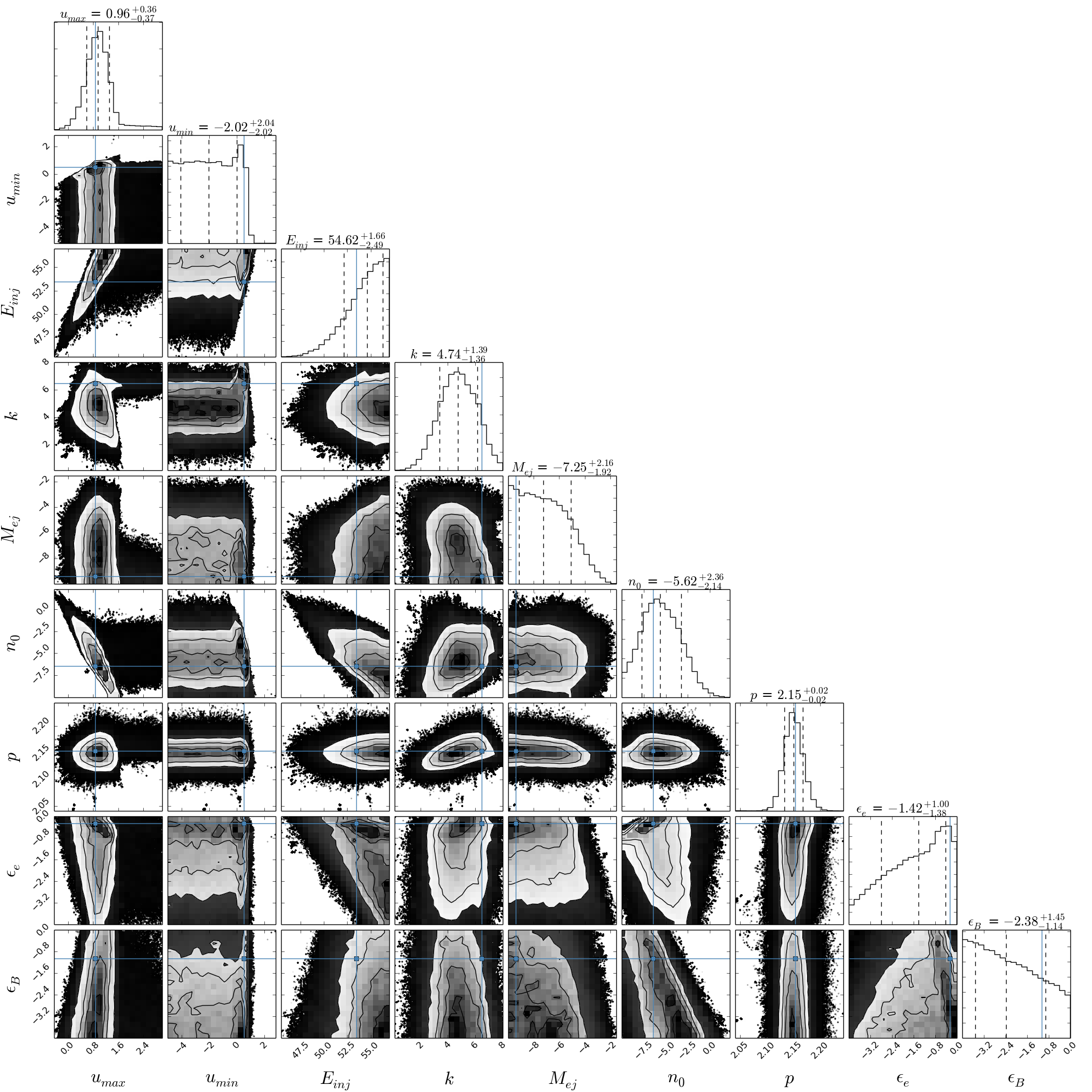}
    \caption{Fit result for the cocoon model. This ``corner plot'' shows all one-dimensional (diagonal) and two-dimensional (off-diagonal) projections of the posterior pdf. 
    The best-fit value (maximum posterior probability) is shown in blue. Dotted lines mark the 16\%, 50\%, and 84\% quantiles of the marginalized posteriors for each parameter. 
    }
    \label{fig:lcjet}
\end{figure*}

%%%%%%%%%%%%%%%%%%%%%%%%%%%%%%%%%%%%%%%%%%%%%%%%%%

%%%%%%%%%%%%%%%%% APPENDICES %%%%%%%%%%%%%%%%%%%%%

%\appendix

%\section{Structured Jet Description}
%\label{app:jet}

%Jet jet jet.

%\section{Cocoon Description}
%\label{app:cocoon}

%If you want to present additional material which would interrupt the flow of the main paper,
%it can be placed in an Appendix which appears after the list of references.

%%%%%%%%%%%%%%%%%%%%%%%%%%%%%%%%%%%%%%%%%%%%%%%%%%

% Don't change these lines
\bsp	% typesetting comment
\label{lastpage}
\end{document}